\documentclass[journal]{IEEEtran}%
\IEEEoverridecommandlockouts
\usepackage{soul}
\usepackage{cite}
\usepackage{amsmath,amssymb,amsfonts}
\usepackage{amsthm,bm}
\usepackage{colortbl}
\usepackage{hhline}
\usepackage{graphicx}
\usepackage{subfigure} 
\usepackage{textcomp}
\usepackage{url}
\usepackage{stfloats}
\usepackage{mathtools}
\usepackage{cuted}
\usepackage{array}  
\usepackage {threeparttable}
\usepackage{tablefootnote}
\usepackage{multirow}
\usepackage{colortbl}
\def\BibTeX{{\rm B\kern-.05em{\sc i\kern-.025em b}\kern-.08em
    T\kern-.1667em\lower.7ex\hbox{E}\kern-.125emX}}

\usepackage{stackengine}
\usepackage{enumerate}
\usepackage{booktabs}
\usepackage{subfig}
\usepackage{multicol}
\usepackage[noend]{algpseudocode}
\usepackage[table,xcdraw]{xcolor}
\usepackage{pgfplots}
\usepackage{pgfplotstable}
\usepackage[ruled,linesnumbered]{algorithm2e} 
\SetKwRepeat{Do}{do}{while}
\definecolor{Reds}{RGB}{0,0,0}
\definecolor{Blues}{RGB}{0,0,0}
\usepackage{setspace}
\usepackage{color}
\usepackage{tabularray}
\definecolor{Iron}{rgb}{0.811,0.815,0.815}
\makeatletter
\def\BState{\State\hskip-\ALG@thistlm}
\makeatother

\algnewcommand\algorithmicforeach{\textbf{for each}}
\algdef{S}[FOR]{ForEach}[1]{\algorithmicforeach\ #1\ \algorithmicdo}
\begin{document}

\title{Semantic Change Driven Generative Semantic Communication Framework

}
\author{Wanting Yang\textsuperscript{a}, Zehui Xiong\textsuperscript{a}, Hongyang Du\textsuperscript{b}, Yanli Yuan\textsuperscript{c}, Tony Q. S. Quek\textsuperscript{a} 
\\
\textsuperscript{a}Information Systems Technology and Design Pillar, Singapore University of Technology and Design\\
\textsuperscript{b} School of Computer Science and Engineering, Nanyang Technological University\\
\textsuperscript{c} School of Cyberspace Science and Technology, Beijing
Institute of Technology\\
\{wanting\_yang, zehui\_xiong, tonyquek\}@sutd.edu.sg,  hongyang001@e.ntu.edu.sg, yanliyuan@bit.edu.cn   

}

\makeatletter
\setlength{\@fptop}{0pt}
\makeatother

\maketitle

\vspace{-1.8cm}
\begin{abstract}
The burgeoning generative artificial intelligence technology offers novel insights into the development of semantic communication (SemCom) frameworks. These frameworks hold the potential to address the challenges associated with   the black-box nature inherent in existing end-to-end training manner for the existing SemCom framework, as well as deterioration of the user experience caused by the inevitable error floor in  deep learning-based SemCom. In this paper, we focus on the widespread remote monitoring scenario, and propose a semantic change driven generative SemCom framework. Therein, the semantic encoder and semantic decoder can be optimized independently. Specifically, we develop a modular semantic encoder with value of information based semantic sampling function. In addition, we propose a conditional
denoising diffusion probabilistic mode-assisted semantic decoder that relies  on received semantic information from the source, namely, the semantic map, and the local static scene information to remotely regenerate scenes. Moreover, we demonstrate the effectiveness of the proposed semantic  encoder and decoder as well as  the considerable potential in
reducing energy consumption through simulation based on the realistic $\mathcal{F}$ composite
channel fading model. 
The code is available at \url{https://github.com/wty2011jl/SCDGSC.git}~\footnote{It will be public when the paper is accepted.}
\end{abstract}


\newtheorem{definition}{Definition}
\newtheorem{lemma}{Proposition}
\newtheorem{theorem}{Theorem}

\newtheorem{property}{Property}

\vspace{0cm}

\section{Introduction}
As the human society progresses toward the remote management and automation, a high-capacity system ensuring reliability and cost/power-efficiency  will emerge as an imperative requirement~\cite{iyer2023survey}.  To achieve optimal effectiveness and the sustainability against this background, semantic communication (SemCom) has garnered considerable attention as a promising solution. Exploiting the intelligence of the communicating parties,  SemCom  shifts the focus from ``how" to transmit to ``what" to transmit~\cite{popovski2020semantic}, so as to boost network performance by reducing the data required to be transmitted.  

Specifically,  the research endeavors in SemCom primarily concentrate on how to extract absolutely required information for recovering the meaning, i.e., the design of the semantic encoder and semantic decoder.
The existing SemCom framework can be broadly divided into three categories based on the adopted semantic extraction methods, i.e., deep learning (DL) based SemCom, reinforcement learning (RL) based SemCom, and knowledge base (KB) assisted SemCom~\cite{yang2022semantic}. Among them, DL-based SemCom has emerged as the prevailing choice for a diverse spectrum of communication tasks~\cite{qin2021semantic}. Nevertheless, the intrinsic limitations of DL give rise to two inevitable challenges with SemCom. Firstly,  the requirement for a differentiable loss function places constraints on the selection of metrics guiding the training process, thereby diminishing its adaptability to a diverse range of tasks. Secondly, the inherent error floor in DL results in sub-optimal communication performance, even under ideal channel conditions. 

To address the first challenge, RL-based SemCom has been proposed to allow for the incorporation of more semantic metrics into the training process~\cite{lu2022rethinking}. However, it should be noted that RL-based SemCom is only applicable to sequence-generation tasks due to the recurrent nature of the actor network.
In addition, both DL-based and RL-based semantic communication exhibit a complete end-to-end black-box nature, which limits these SemCom frameworks’ social acceptance and practicality in complex network environments\cite{yang2022semantic}. 
In comparison, KB-based SemCom excels in terms of explainability. Nevertheless, due to the high computational complexity involved in constructing the KB itself, the application of KB-assisted SemCom to real-time, on-demand tasks is challenging~\cite{yang2023task}.

To address the aforementioned issues, generative artificial intelligence,  particularly the recently prominent conditional denoising diffusion probabilistic  model (DDPM), offers novel insights for the advancement of semantic encoders and decoders. In real life, most communication scenarios do not necessitate perfect recovery at the bit or pixel level; rather, they require the retrieved data to closely approximate reality while ensuring the preservation of complete semantic information. This ensures a favorable quality of experience (QoE). For example, in traffic flow monitoring or parking space surveillance, the system remains indifferent to particulars like the color of the vehicle. Instead, it exclusively focuses on the vehicle's location, necessitating a high level of image clarity.    To this end, we introduce the conditional DDPM into the SemCom framework, where the sender only needs to transmit essential semantic information, expertly extracted by a customised semantic encoder, as a prompt to the receiver with the DDPM assisted semantic decoder. The receiver, in turn, utilizes this prompt to steer purposeful generation to fulfil the task of semantic decoding. Thanks to the noteworthy accomplishments in a plethora of real-world generation
tasks, especially the photo-realistic
images generation, the DDPM model based semantic decoder exhibits the capability to alleviate the deterioration  of QoE  resulting from inevitable error floors inherent in DL-based SemCom framework.

Taking the above into consideration, we propose a novel generative Semcom framework explicitly tailored for the widespread remote monitoring scenario for the first time, called semantic change driven generative semantic
communication (SCDGSC) framework. In contrast to the three  prevailing SemCom frameworks, all of which adhere to the end-to-end training paradigm, the generative SemCom distinguishes itself by affording the opportunity for independent design and optimization of the semantic encoder and semantic decoder, thereby enhancing the explainability of semantic information.
The specific contributions are as follows. 

\begin{itemize}
    \item We have developed a modular semantic encoder endowed with semantic sampling capabilities. In this design, we introduce a more sophisticated  semantic criterion for sampling beyond age, which we  refer to as  value of information (VoI). Given that the primary concern of the receivers predominantly revolves around changes that exert influence on subsequent tasks, the measurement of VoI  concurrently encompasses  the semantic change degree of the observed scene and the age of information (AoI), where the changes irrelevant to the task are omitted.
    \item We have designed a DDPM-assisted semantic decoder, which exclusively relies on the semantic information conveyed by the source, specifically, the semantic map, for the purpose of remote scene generation. Moreover, in order to generate a close-to-real remote scene, an image of static information of the remote scene is also used as one of the inputs to the semantic decoder. Since it is only updated by the source to the destination when static information alterations, such as changes in weather conditions or the time of day, the corresponding communication overhead is negligible.
    \item We have conducted training on the models of the target segmentation module and the DDPM-based scene generation module within semantic encoder and semantic decoder, respectively. The adopted dataset is  generated from CDnet2014~\cite{wang2014cdnet}. Subsequently, we have evaluated the efficiency of these models and have provided substantial evidence of the framework's considerable potential in reducing energy consumption through simulation based on the realistic $\mathcal{F}$ composite
channel fading model.
\end{itemize}

\section{System Model}
In this work, we focus on a single-source and single-server remote status update system. Specifically, the source in the considered scenario is an embedded vision sensor with limited memory and computing power, which is responsible for monitoring a certain scene, generating a series of  visual samples, and updating the destination on the sampled image timely via the wireless and wired transmission. The destination is a remote server, which is  for reproducing the remote monitoring scene in real time for situational awareness, location tracking, control etc.~\cite{9919752}, based on the received samples and the built-in predictive estimation algorithm, such as Kalman Filter and future frame prediction.

In sharp contrast to the conventional communication dedicated to the optimization of transmission process, (where the sampled data flow is usually modelled as a stationary stochastic process or it is assumed that the source adopts periodic sampling), in this work, the optimization of sampling process is also factored into the communication process. The embedded source can semantically sample and extract most pivotal information to promise a significant reduction in transmission burden, thus saving considerable resources while guaranteeing communications performance. The details of the implementation for the proposed tailored SemCom framework are presented and studied in Section~\ref{sec:SCGSC}.

In addition,  we believe that  wireless transmission is a bottleneck in the communication process. In this sense, we mainly analyze and evaluate the wireless transmission performance in this work. 
Considering the combined effects of multi-path and shadowing on the practical transmission, we adopt the $\mathcal{F}$ composite
fading model to characterize the stochastic wireless channel~\cite{8638956}. We denote
the instantaneous channel gain by  $\tilde g$. The probability density function of  $\tilde g$ is expressed by~\cite{8638956}
\begin{equation}
    f\left( {\tilde g} \right) = \frac{{{m^m}{{\left( {{m_s} - 1} \right)}^{{m_s}}}{{\bar g}^{{m_s}}}{{\tilde g}^{m - 1}}}}{{B\left( {m,{m_s}} \right){{\left[ {m\tilde g + \left( {{m_s} - 1} \right)\bar g} \right]}^{m + {m_s}}}}},\label{pdf}
\end{equation}
where $m$, $m_s$ represents the number of clusters of multipath, shadowing shape, respectively, and ${\bar g}$ is corresponding average channel gain, i.e., $\bar g = \mathbb{E}\left[ {\tilde g} \right]$. Moreover, $B\left( { \cdot , \cdot } \right)$ denotes the beta function~\cite{8638956}. 
In this work, we focus on the design of the semantic encoder and decoder. Without loss of generality, we assume perfect
capacity achieving coding in this work. It is assumed that in order to cope with stochastic fading, the transmitter of the embedded  vision sensor adopts a power control technique. We denote the decoding threshold of signal-to-noise ratio by $\Theta $.
In this sense, the achievable transmission rate is expressed by
\begin{equation}
    {R} = W\log \left( {1 + \Theta } \right), 
\end{equation} where $W$ is the allocated bandwidth. Then,
 the instantaneous transmit power, then, is expressed as
\begin{equation}
\tilde p = \frac{{\Theta {\sigma ^2}}}{{\tilde g}}.
\label{P}
\end{equation}
 Since the stochastic fading can be treated as independently and identically
distributed (i.i.d.) among transmission time intervals, the average transmit power over the time can be expressed by 
\begin{equation}
  \begin{aligned}
     &{{\bar p}} = \mathbb{E}{_{\tilde g}}\left[ {{ \tilde p}} \right] = \int_0^\infty  {\frac{{{\Theta}{\sigma ^2}}}{{\tilde g}}f\left( {\tilde g} \right)d\tilde g}\\
     &={{\Theta}{\sigma ^2}}\int_0^\infty  {{{\tilde g}^{ - 1}}f\left( {\tilde g} \right)d\tilde g}  \\
     & = {{\Theta}{\sigma ^2}}{\mathbb{E}}\left[ {{{\tilde g}^{ - 1}}} \right],\label{sm}
\end{aligned}  
\end{equation}
According to \eqref{pdf}, with the aid of \cite[eq. (3.194.3)]{zwillinger2007table}, the ${n^{{\rm{th}}}}$ moment of ${\tilde g}$ can be derived as 
\begin{equation}
{\mathbb{E}}\left[ {{{\tilde g}^n}} \right] = \frac{{{{\left( {{m_s} - 1} \right)}^n}{{\bar g}^n}\Gamma \left( {m + n} \right)\Gamma \left( {{m_s} - n} \right)}}{{{m^n}\Gamma \left( m \right)\Gamma \left( {{m_s}} \right)}},  \label{moment}
\end{equation}
 where  $\Gamma \left(  \cdot  \right)$ represents the gamma function. Substituting the case of $n=-1$ in \eqref{moment} into \eqref{sm}, we can obtain the final expression of  ${{\bar p}}$ as below,
\begin{equation}
    {{\bar p}} = \frac{{{\Theta}{\sigma ^2}}{m\Gamma \left( {m - 1} \right)\Gamma \left( {{m_s} + 1} \right)}}{{\left( {{m_s} - 1} \right)\bar g\Gamma \left( m \right)\Gamma \left( {{m_s}} \right)}}. \label{ap}
\end{equation}

\label{Description}
\begin{figure*}[t]
 \centering
\includegraphics[scale = 0.55]{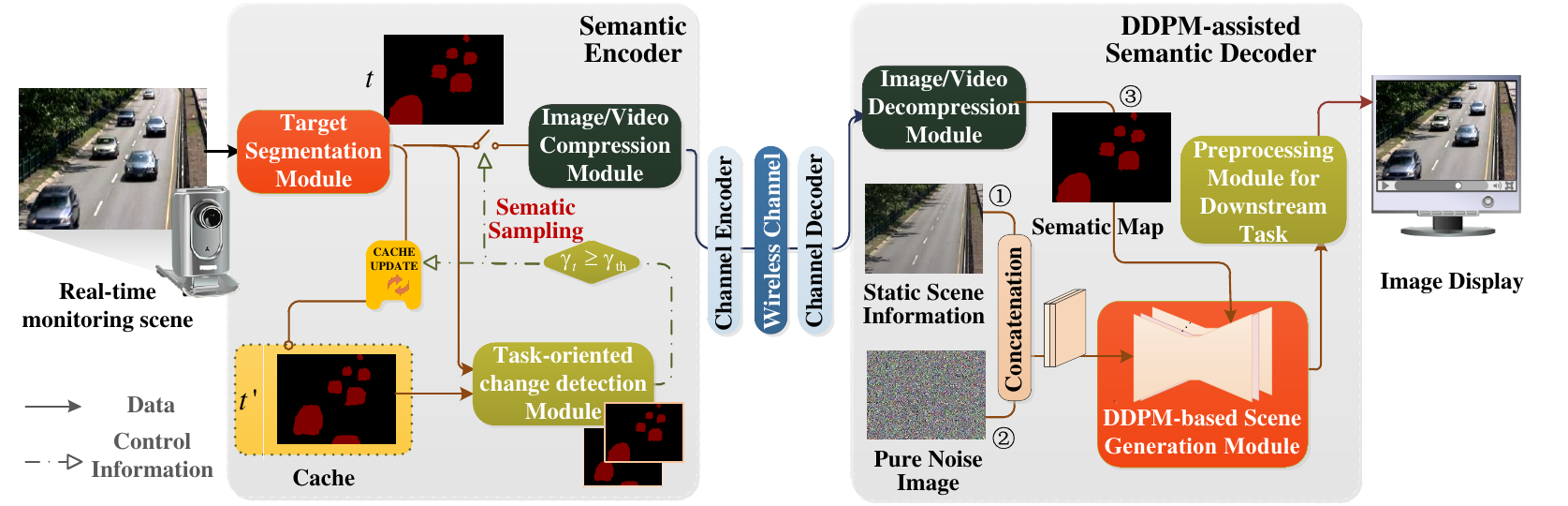}\\
 \caption{Semantic change driven generative SemCom framework. 
 }
\label{fig:Framework}
\vspace{-0.3cm}
\end{figure*}

\section{SCDGSC: Framework and Key Components}
\label{sec:SCGSC}
\subsection{Framework Overview}
\label{sec: overview}
In this work, we develop a novel SCDGSC framework, as illustrated in Fig.~\ref{fig:Framework}. 
In contrast to the existing SemCom model characterized by black-box nature of the end-to-end training manner, a divide-and-conquer approach is adopted in the proposed framework. Moreover, compared to the source encoder in the conventional communication, which only serves the role of video or image compression, two additional modules, \textit{target segmentation} and \textit{VoI-based semantic sampling}, are integrated in the deigned semantic encoder. Accordingly, the semantic decoder is also integrated two modules, \textit{DDPM-based scene generation} and \textit{preprocessing
module for
downstream
task}, which correspond to target segmentation module and VoI-based sampling module,  respectively.

It is widely recognized that, in the remote monitoring system, only the information about changes is of concern, especially the changes in mission-related objectives. With this in mind, the captured scene is first fed into the target segmentation module to extract the location and contour information of the objectives of interest and simultaneously timestamped with $t$. The output of the target segmentation module is hereinafter referred to as the \textit{semantic map}, which is denoted by ${\bf{s}}$. For tracking changes in the scene, the source maintains a cache to store the last semantic map updated to the destination,  the timestamp of which is denoted by $t'$. To judge the value of the newly captured scene, both the semantic maps ${\bf{s}}_t$ and ${\bf{s}}_{t'}$ are fed into the VoI-based semantic sampling module. Different from the AoI-oriented real-time tracking systems, the module for semantic sampling not only takes the age $\left| {t - t'} \right|$ into account, but the semantic changes in the observed scene. We assume that a VoI threshold (denoted by ${\gamma _{\rm th}}$) is set in the system.  If ${\gamma _t}$ is less than ${\gamma _{\rm{th}}}$, the current semantic map ${\bf{s}}_t$ is discarded directly at the source. 
Otherwise, the semantic map ${\bf{s}}_{t'}$ is replaced by ${\bf{s}}_t$ for the next sampling judgement. At the same time, the semantic map ${\bf{s}}_t$ is fed into the compression model and then sent to the destination as an update sample.

 As opposed to the process of semantic encoding, the received compressed updated samples after the channel decoder are firstly fed into the image/video decompression module. Then, the recovered semantic maps are taken as one of the three inputs of DDPM-based scene generation module to guide the  generation of the real-time remote scene. Moreover, for synthesizing a seamless and realistic scene, the static scene information of the remote scene also acts as an input of the generation module, which is denoted by~$\bf r$. In addition, the third input is a pure noise image, which is the outcome of the Gaussian-based forward process of DDPM model itself. At last, the generated scene images are input into the preprocesing module for the downstream task. Take video surveillance reconstruction as an example. The generated images  as well as their timestamp information can be input into a  frame prediction module, which can present users with the illusion of the real-time transmission for the remote scene, by leveraging the frame prediction techniques. The details of the key component design can be found in the next subsection.

\subsection{Key Component Design}
Given the space limitation, only three modules proposed in this work, \textit{target segmentation}, \textit{VoI-based semantic sample}, and \textit{DDPM-based scene generation}, are presented here for the implementation process. 

\subsubsection{Target Segmentation}
The design of the target segmentation can be divided into four parts. The first block is the initial block, which is mainly used to downsample the captured image of the scene so as to reduce the size of the subsequent feature map. The second part is the backbone, which is employed to extract the semantic information embedded in the original image. Given the limited computing capacity of the embedded  vision sensor,  the backbone of MobileNetV3~\cite{howard2019searching} is chosen in this work for it superior balance between computational efficiency and accuracy. During the semantic extraction, four feature maps can be acquired, two of which are fed into the segmentation head part for the further semantic aggregation and the final target segmentation results. Moreover, to ensure the transmission quality, we add a channel adaptive interpolate block to resize the semantic map before output, which is controlled by a downsampling parameter. The specific mapping of downsampling parameters to channel conditions needs further sutdies in particular scenarios.

\subsubsection{VoI-based Semantic Sampling Module}
The VoI metric considered in this work encompasses two aspects. One is the age of the sample updated to the destination, which can be approximated by calculating the difference between the timestamps of the current captured scene and the last updated scene. As stated in Section~\ref{sec: overview},  it can be expressed by
\begin{equation}
    {\gamma_t ^{\rm AoI}} = \left| {t - t'} \right|.
\end{equation}
The other is the semantic change degree, which can be obtained by comparing the differences between two semantic maps ${\bf{s}}_t$ and ${\bf{s}}_{t'}$. Since the semantic map contains only the location and contour information of the target of interest to the task, the changes about the irrelevant information, such as times of the day and weather, are self-ignored during the comparison. We denote the total number of the pixels occupied by the task-relevant objectives in semantic maps ${\bf{s}}_t$ and ${\bf{s}}_t'$ by $n_t$ and $n_{t'}$, respectively. Moreover, the number of the pixels in the intersection of the  pixel set occupied by the objectives in the two semantic maps is denoted by $n_{tt'}$. Then, the semantic change degree can be expressed by 
\begin{equation}
    {\gamma_t^{{\rm{change}}}} = \frac{{{n_t} + {n_{t'}} - 2{n_{tt'}}}}{{{n_t} + {n_{t'}}}}. \label{eq:change}
\end{equation}
From \eqref{eq:change}, we can see that, if the target regions in the two semantic maps overlap exactly, ${\gamma _{{\rm{change}}}} = 0$, which means that the scene has not undergone a semantic change in the meantime. In contrast, if the target regions in the two semantic maps are completely separated,  ${\gamma _{{\rm{change}}}} = 1$, which is the maximum value of the semantic change degree. Since in real-world scenarios, the generation at the destination side needs to rely on predictive techniques, the  information that the scene has not changed can also facilitate a better grasp of the evolution of the environment. With this in mind, considering the both factors, the complete expression of VoI can be formulated as
\begin{equation}
    {\gamma _t} = {\tau _1}\gamma _t^{{\text{AoI}}} + {\tau _2}\gamma _t^{{\text{change}}},
\end{equation}
 where parameters $\tau_1$ and $\tau_2$ are used to adjust the sensitivity of the VoI to the semantic change degree and AoI. 
\subsubsection{DDPM-based Scene Generation Model}
Inspired of the remarkable success of the DDPM model in a plethora of real-world generation
tasks, we involve the conditional DDPM into the SemCom framework as the core of semantic decoder. 
As discussed in Section~\ref{sec: overview}, both static scene $\bf r$ and the real-time semantic map ${\bf s}_t$  are taken as the inputs to guide the generation process. Therefore, the conditional DDPM can be treated as a latent variable model of the form ${p_\theta }\left( {{{\bf{x}}_0}} \right): = \int {{p_\theta }\left( {{{\bf{x}}_{0:N}}\left| {{\bf r},{{\bf s}_t}} \right.} \right)d{{\bf{x}}_{0:N}}} $, where
${{\bf x}_1}, \ldots ,{{\bf x}_T}$ are latents with the same dimensionality as the possible generated data ${{\bf x}_0} \sim q\left( {{{\bf x}_0}} \right)$, and ${p_\theta }\left( {{{\bf{x}}_{0:N}}\left| {{\bf{r}},{{\bf{s}}_t}} \right.} \right) = p\left( {{{\bf{x}}_N}} \right)\prod\nolimits_{n = 1}^N {{p_\theta }\left( {{{\bf{x}}_{n - 1}}\left| {{{\bf{x}}_n},{\bf{r}},{{\bf{s}}_t}} \right.} \right)}$. The joint distribution
${{p_\theta }\left( {{{\bf{x}}_{0:N}}\left| {{\bf r},{{\bf s}_t}} \right.} \right)}$ is called the reverse process, which can be modelled as a Markov chain with learned Gaussian
transitions 
\begin{equation}
 {p_\theta }\left( {{{\bf{x}}_{n - 1}}\left| {{{\bf{x}}_n},{\bf{r}},{{\bf{s}}_t}} \right.} \right) = {\cal N}\left( {{{\bf{x}}_{n - 1}};{{{\bf{\tilde \mu }}}_\theta }\left( {{{\bf{x}}_n},{{\bf{x}}_0}} \right),{{\tilde \beta }_\theta }I} \right)    
\end{equation}
starting at a pure noise image ${{\bf x}_N} \sim {\cal N}\left( {{{\bf x}_N}; {\bf 0},{\rm \bf I}} \right)$. To facilitate learning ${{p_\theta }\left( {{{\bf{x}}_{0:N}}\left| {{\bf r},{ {\bf s}_t}} \right.} \right)}$, an approximate posterior $q\left( {{{\bf x}_{1:N}}\left| {{{\bf x}_0}} \right.} \right)$ named forward process is defined and fixed to a Markov chain that
progressively adds Gaussian noise into the data under a variance schedule ${\beta _1}, \ldots ,{\beta _N}$\footnote{In this work, the variance schedule is held constant as hyperparameters}. Given that ${\bar \alpha _n}: = \prod\nolimits_{s = 1}^n {\left( {1 - {\beta _s}} \right)}$, the forward process is expressed by 
\begin{equation}
    q\left( {{{\bf{x}}_n}\left| {{{\bf{x}}_{n - 1}}} \right.} \right) = {\cal N}\left( {{{\bf{x}}_n};\sqrt {1 - {\beta _n}} {{\bf{x}}_0},{\beta _n}{\bf I}} \right), \label{eq:f1}
\end{equation}
\begin{equation}
    q\left( {{{\bf x}_n}\left| {{{\bf x}_0}} \right.} \right) = {\cal N}\left( {{{\bf x}_n};\sqrt {{{\bar \alpha }_n}} {{\bf x}_0},\left( {1 - {{\bar \alpha }_n}} \right){\bf I}} \right).\label{eq:f2}
\end{equation}
Moreover, following the properties of the Gaussian distribution, the posteriors of the forward process, $q\left( {{{\bf{x}}_{n - 1}}\left| {{{\bf{x}}_n},{{\bf{x}}_0}} \right.} \right)$, also obeys a Gaussian distribution,i.e.,
\begin{equation}
q\left( {{{\bf{x}}_{n - 1}}\left| {{{\bf{x}}_n},{{\bf{x}}_0}} \right.} \right) = {\cal N}\left( {{{\bf{x}}_{n - 1}};{\bf{\tilde \mu }}_n\left( {{{\bf{x}}_n},{{\bf{x}}_0}} \right),{{\tilde \beta }_n}{\bf I}} \right).
\end{equation}
Based on \eqref{eq:f1} and \eqref{eq:f2}, we have~\cite{ho2020denoising}
\begin{equation}
    {{\tilde \mu }_n}\left( {{x_n},{x_0}} \right) = \frac{1}{{\sqrt {1 - {\beta _n}} }}{{\bf{x}}_n} - \frac{{{\beta _n}}}{{\sqrt {1 - {\beta _n}} \sqrt {1 - {{\bar \alpha }_n}} }}\epsilon, \label{eq:miu}
\end{equation}
\begin{equation}
    {{\tilde \beta }_n} = \frac{{1 - {{\bar \alpha }_{n - 1}}}}{{1 - {{\bar \alpha }_n}}}{\beta _n}. \label{eq:beta}
\end{equation}
\begin{figure}[t]
 \centering
\includegraphics[scale = 0.52]{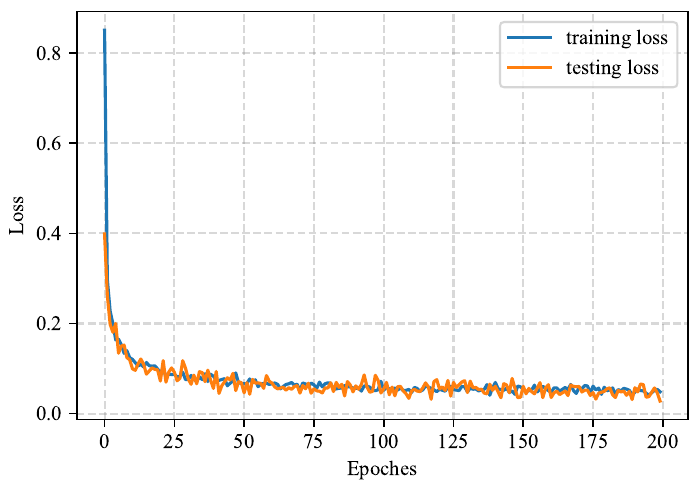}\\
 \caption{Loss function for conditional DDPM.
 }
\label{fig:loss}
\vspace{-0.5cm}

\end{figure}
\begin{table}[b]
\centering
 \caption{
 Performance of target segmentation.}
\begin{tabular}{c|c} 
\hline
Metrics                       & Value              \\ 
\hline
Average
row correct           & {[}`99.5',`71.9']  \\
Intersection over Union (IoU) & {[}`97.9',`65.6']  \\
mean IoU                      & 81.7               \\
\hline
\end{tabular} \label{Tab:peformance}
\end{table}
The conditional DDPM is trained to optimize the upper variational bound on negative log likelihood via minimizing the gap between $q\left( {{{\bf{x}}_{n - 1}}\left| {{{\bf{x}}_n},{{\bf{x}}_0}} \right.} \right)$ and ${{p_\theta }\left( {{{\bf{x}}_{n - 1}}\left| {{{\bf{x}}_n},{\bf{r}},{{\bf{s}}_t}} \right.} \right)}$. According to \eqref{eq:beta}, in our work, the coefficient of variance can be considered as a constant. As such, in this DDPM, it is only required to make ${{{{\bf{\tilde \mu }}}_\theta }\left( {{{\bf{x}}_n},{{\bf{x}}_0}} \right)}$ as close as possible to ${{{{\bf{\tilde \mu }}}_n}\left( {{{\bf{x}}_n},{{\bf{x}}_0}} \right)}$. 
By the observation of \eqref{eq:miu}, an U-Net network ${{\hat \epsilon}_\theta }\left( {{{\bf{x}}_n},{\bf{r}},{{\bf{s}}_t}} \right)$ is employed to approximate the noise $\epsilon$ generated in each step.  

Following the formulation of~\cite{ho2020denoising}, the simplified denoising  loss function  can be expressed by 
\begin{equation}
    {\mathcal{L}_d} = {\mathbb{E}_{n,{{\mathbf{x}}_n},\epsilon}}\left[ {{{\left\| {\epsilon - {\epsilon_\theta }\left( {\sqrt {{{\bar \alpha }_n}} {{\mathbf{x}}_n} + \sqrt {1 - {{\bar \alpha }_n}} \epsilon,{\mathbf{r}},{{\mathbf{s}}_t},n} \right)} \right\|}_2}} \right]
\end{equation}
In addition, according to~\cite{dhariwal2021diffusion}, the performance of conditional DDPM can be enhance by gradient of the log probability distribution ${\nabla _{{{\mathbf{x}}_n}}}\log p\left( {{{\mathbf{s}}_t}\left| {{{\mathbf{x}}_n}} \right.} \right)$. In this work, we adopt the classifier-free guidance to implicitly infers the gradient of the log
probability~\cite{wang2022semantic}, as shown in \eqref{eq:free}. Specifically, the semantic map ${{\mathbf{s}}_t}$ is replaced with a null label $\emptyset$ to disentangle the noise estimated under the guidance of semantic map
${\epsilon_\theta }\left( {{{\mathbf{x}}_n}\left| {{\mathbf{r}},{{\mathbf{s}}_t}} \right.} \right)$  from unconditional situation ${\epsilon_\theta }\left( {{{\mathbf{x}}_n}\left| {\mathbf{r}} \right.} \right)$.

\begin{equation}
\begin{split}
  & {\epsilon_\theta }\left( {{{\mathbf{x}}_n}\left| {{\mathbf{r}},{{\mathbf{s}}_t}} \right.} \right) - {\epsilon_\theta }\left( {{{\mathbf{x}}_n}\left| {\mathbf{r}} \right.} \right) \\ \propto & {\nabla _{{{\mathbf{x}}_n}}}\log p\left( {{{\mathbf{x}}_n}\left| {{{\mathbf{s}}_t},{\mathbf{r}}} \right.} \right) - {\nabla _{{{\mathbf{x}}_n}}}\log p\left( {{{\mathbf{x}}_n}\left| {\mathbf{r}} \right.} \right) \hfill \\
   \propto &{\nabla _{{{\mathbf{x}}_n}}}\log p\left( {{{\mathbf{s}}_t}\left| {{{\mathbf{x}}_n}} \right.} \right) \hfill 
\end{split} \label{eq:free}
\end{equation}
Thus, the noise estimation can be performed based on the disentangled component, which can be refined as 
\begin{equation}
    {{\hat \epsilon}_\theta }\left( {{{\mathbf{x}}_n}\left| {{\mathbf{r}},{{\mathbf{s}}_t}} \right.} \right) = {\epsilon_\theta }\left( {{{\mathbf{x}}_n}\left| {{\mathbf{r}},{{\mathbf{s}}_t}} \right.} \right) + k \cdot \left( {{\epsilon_\theta }\left( {{{\mathbf{x}}_n}\left| {{\mathbf{r}},{{\mathbf{s}}_t}} \right.} \right) - {\epsilon_\theta }\left( {{{\mathbf{x}}_n}\left| {\mathbf{r}} \right.} \right)} \right),
\end{equation}
where $k$ is the guidance scale, which allows the generated data to follow the semantic map more strictly.
\begin{figure*}[t]
 \centering
 \subfigure[]{
\includegraphics[scale = 0.95]{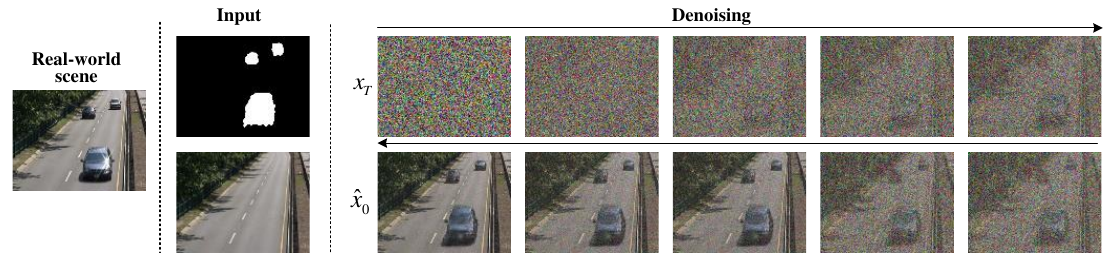}
}\\
 \subfigure[]{
 \includegraphics[scale = 0.95]{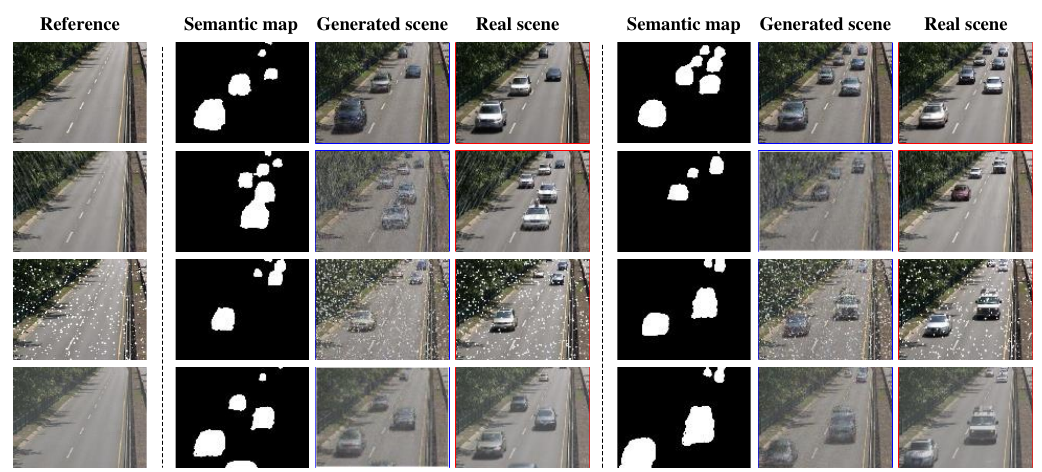}
 }
 \caption{Visual results. (a) Inference procedure; (b)  Performance comparison in different weather conditions. 
 }
\label{fig:compare}
\vspace{-0.3cm}
\end{figure*}
\section{Evaluation}
\label{sec:eva}
\subsection{Setup}
\subsubsection{Datasets}
We select data from the `baseline' category within the CDNet2014 dataset~\cite{wang2014cdnet}, specifically focusing on road traffic scenarios, as our training and evaluation dataset  for target segmentation and conditional DDPM. Specifically, in the training of conditional DDPM, the image of moment``t0" in the CDnet2014 is treated as the reference image and the results of the target segmentation are used as the semantic map. The images at other moments are treated as the labels. We add weather filters for rain, snow and fog to all the images. The training and test sets are divided in a ratio of 8 : 2. The image size is reshaped into $\left( {128,96} \right)$.
\subsubsection{Hyperparameters}
For the target segmentation, we employ the architecture of Lite R-ASPP and trained on the basis of the pre-training weights obtained by pre-training on COCO~\footnote{https://download.pytorch.org\/models\/lraspp\_mobilenet\_v3\_large-d234d4ea.pth}. Moreover, we set the number of class as 2. For the conditional DDPM, the structure of employed U-Net network can be referred to in~\cite{wang2022semantic}. It comprises a number of channels equal to
[64, 64, 128, 128, 256, 256, 512, and 512].
We have $T = 1000$, and a linear variance schedule. 
Finally, the guidance scale $k$ is set to 4. The batch size is set to 6 and the learning rate is set to $2e - 5$.

\subsubsection{Simulation parameter}
For the adopted $\mathcal{F}$ composite fading model, we take fading severity of  $m = 6$, shadowing shape $m_s = 6$. Moreover, the average channel gain is treated as the
 pass loss, which is modeled as $35.3 + 37.6{\log _{10}}(d)$ in dB. The distance between the  vision sensor and the wireless base station is $d = 100$m. The SNR threshold is set to 15~dB. The noising power is -90~dBm/Hz. Moreover, given the scenario we focus on, we set ${\tau _1} = 0$ and ${\tau _2} = 1$.
\subsection{Results Analysis}
\begin{figure}[t]
 \centering
 \subfigure[]{
 \includegraphics[scale = 0.5]{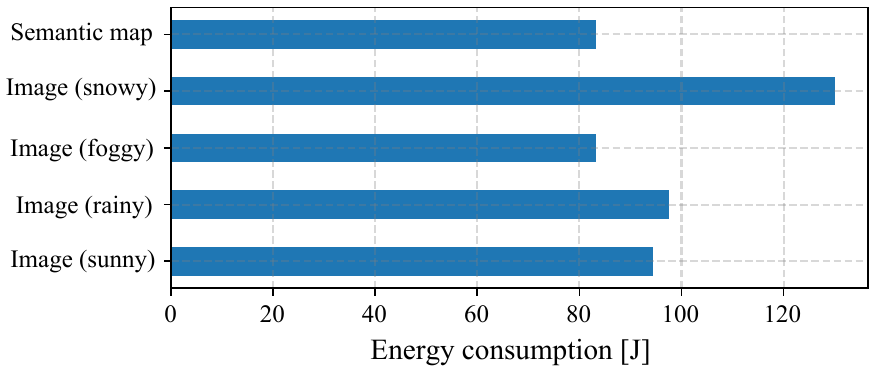}
 }
\\
\subfigure[]{
 \includegraphics[scale = 0.5]{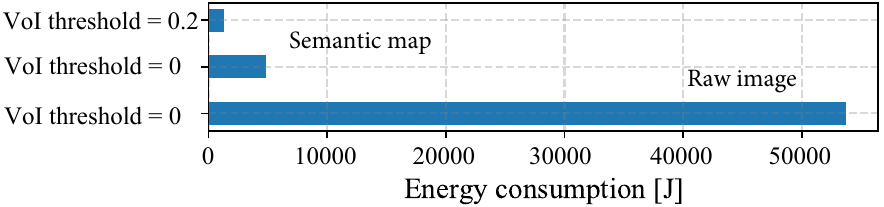}
}
 \caption{Comparison of energy consumption. (a) Comparison  under different weather conditions; (b) Comparison with different VoI thresholds.
 }
\label{fig:energy}

\vspace{-0.3cm}

\end{figure}

Applying the re-trained model of Lite-R-ASPP to the test set, the performance results are summarized in Table~\ref{Tab:peformance}. Taking the output of the Lite-R-ASPP network as one of the inputs of the conditional DDPM, the training and the testing loss of the DDPM is shown in Fig.~\ref{fig:loss}. The  visual results can be found in Fig.~\ref{fig:compare}. Specifically, Fig.~\ref{fig:compare}(a) shows the denoising process. The image in the left column is the real-world scene. The middle column shows the semantic map and the reference image, which can be considered as the prompt for the virtual scene generation. The ten images depicted on the right are captured from the denoising process from step 999 to step 0. By comparing the generated image denoted as $\hat x_0$ and the original image, it becomes evident that, while there exist disparities in the specific details and coloration of the vehicles compared to the real scene, the positional alignment of the vehicles remains consistent. This achievement signifies the system's proficiency in preserving essential semantic information, and also allows the framework to be used in the common scenarios that are not concerned with specific details, such as car park space monitoring, traffic flow monitoring, etc. Additionally, as demonstrated in Fig.~\ref{fig:compare}(b), even in the face of diverse weather conditions, the transmission of a semantic map of the real scene suffices. The receiver side adaptively reconstructs a virtual image at the remote location, leveraging the local reference image as a foundation. We adopt the JPEG image compression technique. The datasize of the image  of the real scene shown in Fig.~\ref{fig:compare}(a) in different weather and the semantic map are 93~kb, 96~kb, 82~kb, 128~kb and 5~kb, respectively. The comparison of energy consumption of the image transmission is shown in Fig.~\ref{fig:energy}(a). In addition, when we set the VoI thresholds to different values, the comparison of the total energy of data transmission when monitoring the same scene is shown in Fig.~\ref{fig:energy}(b). Simulation results unequivocally underscore the substantial potential of the proposed framework in reducing energy consumption, which contributes to the sustainability of embedded  vision sensors.

\section{Conclusion}
In this paper, we focused on the remote monitoring scenario and proposed a semantic change driven generative semantic communication framework. The distinctiveness of the envisaged
framework resided in the design of semantic encoder and decoder, which were bolstered by the advanced semantic segmentation and conditional generative AI techniques, respectively.
In contrast to the existing SemCom model characterized by black-box nature of the end-to-end training manner, a divide-and-conquer approach was adopted in the proposed framework. Simulation results demonstrated the effectiveness of the proposed framework and considerable potential for energy savings. 
In the future, we will systematically investigate semantic sampling optimization within resource constraints and the optimal VoI threshold in this framework, employing diverse performance metrics for downstream tasks.











\bibliographystyle{IEEEtran}
\bibliography{ref}



\end{document}